\documentclass[letterpaper,twocolumn,prl,aps,superscriptaddress,amsmath,amssymb,floatfix]{revtex4-1}

\usepackage{mathptmx}
\usepackage{color}
\usepackage{verbatim}
\usepackage{float}
\usepackage{amsmath}
\usepackage{amssymb}
\usepackage{graphicx}
\usepackage{esint}
\usepackage[unicode=true,
 bookmarks=true,bookmarksnumbered=false,bookmarksopen=false,
 breaklinks=false,pdfborder={0 0 1},backref=false,colorlinks=true]
 {hyperref}
\hypersetup{
 linkcolor=magenta,urlcolor=blue,citecolor=blue,pdfstartview={FitH},hyperfootnotes=false}

\makeatletter

\usepackage{textcomp}
\usepackage{epstopdf}

\pdfpageheight\paperheight
\pdfpagewidth\paperwidth

\@ifundefined{textcolor}{}{%
 \definecolor{BLACK}{gray}{0}
 \definecolor{WHITE}{gray}{1}
 \definecolor{RED}{rgb}{1,0,0}
 \definecolor{GREEN}{rgb}{0,1,0}
 \definecolor{BLUE}{rgb}{0,0,1}
 \definecolor{CYAN}{cmyk}{1,0,0,0}
 \definecolor{MAGENTA}{cmyk}{0,1,0,0}
 \definecolor{YELLOW}{cmyk}{0,0,1,0}
}

\usepackage{xcolor}
\usepackage{soul}
\definecolor{blue}{rgb}{0,0,1}
\definecolor{red}{rgb}{1,0,0}
\definecolor{green}{rgb}{0,1,0}

\@ifundefined{textcolor}{}{%
 \definecolor{BLACK}{gray}{0}
 \definecolor{WHITE}{gray}{1}
 \definecolor{RED}{rgb}{1,0,0}
 \definecolor{GREEN}{rgb}{0,1,0}
 \definecolor{BLUE}{rgb}{0,0,1}
 \definecolor{CYAN}{cmyk}{1,0,0,0}
 \definecolor{MAGENTA}{cmyk}{0,1,0,0}
 \definecolor{YELLOW}{cmyk}{0,0,1,0}
}

\usepackage{xcolor}\usepackage{soul}
\definecolor{blue}{rgb}{0,0,1}
\definecolor{red}{rgb}{1,0,0}
\definecolor{green}{rgb}{0,1,0}

\usepackage{soul}

\usepackage{graphicx}
\usepackage{siunitx}

\makeatother

\begin{document}


\title{On-chip 7~GHz acousto-optic modulators for visible wavelengths}

\author{Ji-Zhe Zhang}
\affiliation{CAS Key Laboratory of Quantum Information, University of Science and Technology of China, Hefei 230026, China\\}
\affiliation{Anhui Province Key Laboratory of Quantum Network, University of Science and Technology of China, Hefei 230026, China.}

\author{Yu Zeng}
\affiliation{CAS Key Laboratory of Quantum Information, University of Science and Technology of China, Hefei 230026, China\\}
\affiliation{Anhui Province Key Laboratory of Quantum Network, University of Science and Technology of China, Hefei 230026, China.}

\author{Qing Qin}
\affiliation{CAS Key Laboratory of Quantum Information, University of Science and Technology of China, Hefei 230026, China\\}
\affiliation{Fujian Provincial Key Laboratory of Quantum Manipulation and New Energy Materials, College of Physics and Energy, Fujian Normal University, Fuzhou 350117, China.}

\author{Yuan-Hao Yang}
\affiliation{CAS Key Laboratory of Quantum Information, University of Science and Technology of China, Hefei 230026, China\\}
\affiliation{Anhui Province Key Laboratory of Quantum Network, University of Science and Technology of China, Hefei 230026, China.}

\author{Zheng-Hui Tian}
\affiliation{CAS Key Laboratory of Quantum Information, University of Science and Technology of China, Hefei 230026, China\\}
\affiliation{Anhui Province Key Laboratory of Quantum Network, University of Science and Technology of China, Hefei 230026, China.}

\author{Jia-Qi Wang}
\affiliation{CAS Key Laboratory of Quantum Information, University of Science and Technology of China, Hefei 230026, China\\}
\affiliation{Anhui Province Key Laboratory of Quantum Network, University of Science and Technology of China, Hefei 230026, China.}

\author{Chun-Hua Dong}
\affiliation{CAS Key Laboratory of Quantum Information, University of Science and Technology of China, Hefei 230026, China\\}
\affiliation{Anhui Province Key Laboratory of Quantum Network, University of Science and Technology of China, Hefei 230026, China.}

\author{Xin-Biao Xu}
\email{xbxuphys@ustc.edu.cn}
\affiliation{CAS Key Laboratory of Quantum Information, University of Science and Technology of China, Hefei 230026, China\\}
\affiliation{Anhui Province Key Laboratory of Quantum Network, University of Science and Technology of China, Hefei 230026, China.}

\author{Ming-Yong Ye}
\affiliation{Fujian Provincial Key Laboratory of Quantum Manipulation and New Energy Materials, College of Physics and Energy, Fujian Normal University, Fuzhou 350117, China.}

\author{Guang-Can Guo}
\affiliation{CAS Key Laboratory of Quantum Information, University of Science and Technology of China, Hefei 230026, China\\}
\affiliation{Anhui Province Key Laboratory of Quantum Network, University of Science and Technology of China, Hefei 230026, China.}
\affiliation{CAS Center For Excellence in Quantum Information and Quantum Physics, University of Science and Technology of China, Hefei 230026, China.}

\author{Chang-Ling Zou}
\email{clzou321@ustc.edu.cn}
\affiliation{CAS Key Laboratory of Quantum Information, University of Science and Technology of China, Hefei 230026, China\\}
\affiliation{Anhui Province Key Laboratory of Quantum Network, University of Science and Technology of China, Hefei 230026, China.}
\affiliation{CAS Center For Excellence in Quantum Information and Quantum Physics, University of Science and Technology of China, Hefei 230026, China.}




\date{\today}

\begin{abstract}
A chip-integrated acousto-optic phase modulator tailored for visible optical wavelengths has been developed. Utilizing the lithium niobate on sapphire platform, the modulator employs a 7 GHz surface acoustic wave, excited by an interdigital transducer and aligned perpendicular to the waveguide. This design achieves efficient phase modulation of visible light within a compact device length of merely 200 microns, while holds the advantages of easy fabrication and high stability due to simple unsuspended structure. Remarkably, in this high-frequency acoustic regime, the acoustic wavelength becomes comparable to the optical wavelength, resulting in a notable single-sideband modulation behavior. This observation underscores the phase delay effects in the acousto-optics interactions, and opens up new aspects for realizing functional visible photonic devices and its integration with atom- and ion-based quantum platforms.
\end{abstract}

\maketitle


\section{\label{sec:level1}Introduction}
In the last few decades, the need for efficient on-chip control of visible light has been growing~\cite{Sabouri2018,Mashayekh2021,Chauhan2021,lin2022,Wang2023,Poon2024,Buzaverov2024}. Among various photonic devices operating at visible wavelengths, optical modulators are essential for controlling the frequency and amplitude of light~\cite{Desiatov2019,Liang2021,Celik2022,Dong2022,Park2024}. In particular, the platforms for hybrid  photonic-atomic chip~\cite{Meng2015,Pelucchi2022,Chen2022MOT,liu2022,liu2023,Zhou2024} and the integration of photonic circuits with trapped ions~\cite{Mehta2020,Romaszko2020,Hogle2023} demand coherent control and manipulation of atomic quantum states via visible lasers. For instance, in chip-scale cold atomic clocks and sensors, visible-light modulators can be used to generate sidebands for laser cooling and state preparation of trapped atoms~\cite{Isichenko2023,Kodigala2024,Park2024}. In quantum computing with trapped ions, the modulators enable the precise control of ion qubits through Raman sideband transitions~\cite{Hogle2023,Kwon2024,Park2024}. Beyond quantum applications, visible-light modulators can also find use in bio-photonic sensing~\cite{Mashayekh2021} and chip-scale light detection and ranging (LiDAR) systems for 3D imaging and ranging~\cite{Li2023}.

Along with the development of photonic integrated circuits for telecommunications, various approaches have been explored for on-chip optical modulation, including electro-optic, thermo-optic, and acousto-optic techniques. Electro-optic modulators based on lithium niobate (LN) have shown impressive performance due to LN's excellent electro-optic properties~\cite{Zhu2021,Xie2024,Boes2023,Feng2024}. However, they often suffer from large device footprints, as the weak electro-optic interaction necessitates longer interaction lengths. In particular, the generation of advanced modulation functions, such as single-sideband modulation, typically requires complex device architectures and precise phase control~\cite{Xu2020,Porzi2022,Idres2023,Kodigala2024}. Thermo-optic modulators offer compactness but are limited by their low modulation speed~\cite{Liu2022Rev}. In contrast, acousto-optic modulation exploits the interaction between acoustic and optical waves, offering stronger interactions and the ability to generate frequency-shifted optical signals. Previous works have demonstrated on-chip acousto-optic modulators (AOMs) for telecom photons and low microwave frequencies below 5 GHz, and have been applied in phase and intensity modulations~\cite{Cai2019,Sarabalis2020,Wan2022,Sarabalis2021,huang2022acousto,yu2020acousto}, frequency shifting~\cite{shao2020integrated,Yu2021}, isolators~\cite{Dostart2020,Kittlaus2018,kittlaus2021electrically,Zhang2024}, LiDAR~\cite{Li2023}, and show potential for microwave photonics~\cite{Gertler2022,Yu2024} and optical-microwave conversion~\cite{Shao2019,Hassanien2021,Han2021}.

In this work, we realize on-chip AOMs operating at a record-high frequency of 7~GHz for visible wavelengths. Our device leverages the unique advantages of the lithium niobate on sapphire (LNOS) platform~\cite{Sarabalis2020,Mishra2021,Mayor2021,Shao2022,Yang2023,qinqing}, which provides low acoustic loss at high frequencies and effectively confines the acoustic energy in the slab. By optimizing the interdigital transducer (IDT) design, we achieve efficient acoustic wave excitation, enabling strong acousto-optic interactions within a compact device footprint of merely 200 microns. Remarkably, we observe an unexpected single-sideband modulation behavior in this high-frequency acoustic regime, where the acoustic wavelength becomes comparable to the optical wavelength. This observation reveals potential new physics of acousto-optics interaction at the sub-micron scale and paves the way for novel integrated photonic devices with advanced modulation capabilities.

\section{\label{sec:level2}Device}
\begin{figure}[ht]
	\centering
	\includegraphics[width=1\linewidth]{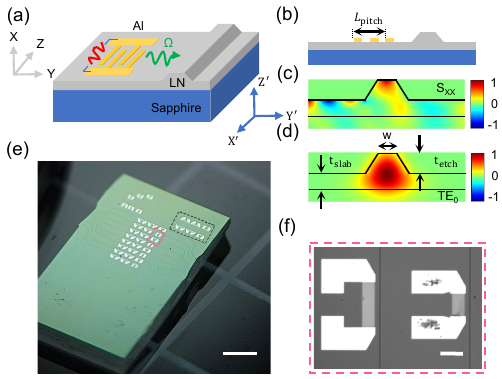}
	\caption{(a) Schematic diagram of the integrated AOM on the thin-film lithium niobate platform. The substrate is sapphire, and the metal electrodes are made of aluminium. X, Y, Z represents the crystal axes of lithium niobate. $\mathrm{X^{'}}$, $\mathrm{Y^{'}}$, $\mathrm{Z^{'}}$ represent the crystal axes of thin-film sapphire. (b) 2D cross-section of the integrated AOM, illustrating the electrode fingers of the IDT for exciting SAW, with the period of the electrode fingers is $L_{\mathrm{pitch}}$. (c) The distributions of strain field ($S_{\mathrm{XX}}$) induced by the SAW. (d) The electric field profiles of fundamental transverse-electric mode at the cross-section of the waveguide. (e) Photograph of the fabricated AOMs on the chip, featuring array of IDTs. (f) Optical micrograph of the AOM device, highlighting the IDT with electrode fingers parallel to the optical waveguide.}
	\label{fig1}
\end{figure}

Fig.~\ref{fig1}(a) presents a schematic diagram of the integrated AOM for visible light, highlighting its key components: the IDT (yellow rectangle array) and the optical wedge waveguide (grey trapezoid). The IDT, consisting of periodic aluminum electrodes, efficiently converts the input radio frequency (RF) signal into surface acoustic waves (SAWs) by harnessing the piezoelectricity of LN. Fig.~\ref{fig1}(b) illustrates the two-dimensional cross-section of the AOM. Through the photoelastic effect, the SAW-induced strains modulate the refractive index of the LN wedge waveguide. The IDT electrodes are aligned parallel to the optical waveguide, ensuring that the excited SAW propagates perpendicular to the waveguide, resulting in a nearly uniform refractive index modulation along the propagation direction of optical wave. Consequently, the accumulated phase of light passing through the SAW interaction region leads to RF-controlled phase modulation. 

We realize the AOM using the LNOS platform~\cite{Sarabalis2020,Mishra2021,Mayor2021,Shao2022,Yang2023,qinqing}, leveraging the strong piezoelectric and photoelastic properties of thin-film LN. The LNOS platform offers several advantages for high-performance AOMs. First, LN's high piezoelectric coefficient enables efficient SAW generation without requiring hetero-integration of other piezoelectric materials, and the strong photoelastic effect is beneficial for acousto-optic interaction. Second, the sapphire substrate provides excellent optical transparency and low optical losses, making it ideal for visible-wavelength integrated photonics. Third, the combination of single-crystal LN and sapphire enables strong SAW confinement in the LN layer, allowing low SAW propagation loss and promising excellent high-frequency SAW devices. Moreover, the LNOS platform enables monolithic integration with other photonic and electronic components, opening possibilities for complex photonic circuits with enhanced functionality and reduced footprint for potential applications~\cite{Xu2022,Okada2021}. Fig.~\ref{fig1}(a) also illustrates the crystal axes of the thin-film LN and the sapphire substrate. The LN is X-cut and the sapphire is Z-cut, with Y-axes of LN and $\mathrm{Y^{'}}$ of sapphire are in-plane and parallel. The SAWs propagate along the Y-axis of the LN, while the optical mode confined in the waveguide propagates along the Z-axis. 

To optimize the AOM performance, we conducted numerical simulations of the coupling between the propagating SAW and optical waveguide mode. As shown in Fig.~\ref{fig1}(b), the IDT has a pitch width ($L_{\mathrm{pitch}}$) equal to four times the width of a single metal electrode, resulting in a $50\%$ duty ratio. Fig.~\ref{fig1}(c$\sim$d) presents the distribution of the strain component ($S_{\mathrm{XX}}$) induced by the SAW and the electric field profile of the optical $\mathrm{TE_{0}}$ mode, with a working RF frequency at $7\,\mathrm{GHz}$ and optical wavelength at $780\,\mathrm{nm}$. According to the numerical simulations based on finite-element method~\cite{qinqing}, we chose a wedge waveguide thickness ($t_{\mathrm{etch}}$) of \SI{230}{\nano\meter} and a thin slab film thickness ($t_{\mathrm{slab}}$) of \SI{170}{\nano\meter} for moderate acousto-optics interaction strength, while maintaining a relative low optical loss. Considering the SAW wavelength on the LN slab is approximately \SI{700}{\nano\meter} at a frequency at $7\,\mathrm{GHz}$, the optical waveguide width ($w$) is set to \SI{320}{\nano\meter} to avoid destructive interference effects of acoustic wave in the wedge waveguide~\cite{qinqing}.

Fig.~\ref{fig1}(e) shows a photograph of the fabricated photonic chip, featuring arrays of AOM devices on a transparent sapphire substrate. Fig.~\ref{fig1}(f) is the enlarged top view, which reveals the device's simple structure: a single straight waveguide and one IDT, with a core device footprint of only \SI{260}{\micro\meter}$\times$\SI{140}{\micro\meter}. The IDT length is \SI{200}{\micro\meter}, comprising 40 electrode pairs. We introduced large electrode pads for RF input via the electric probes. Additionally, we fabricated several pairs of IDT without coupling to the waveguide, as shown by the black rectangle in Fig.~\ref{fig1}(e), to characterize the RF-to-SAW conversion efficiency and bandwidth.

\section{\label{sec:level3}Characterization of IDT}
Efficient conversion of the input RF signal to SAWs is crucial for achieving strong acousto-optic interaction. However, realizing high-frequency IDTs with a working frequency above $5\,\mathrm{GHz}$ presents several practical challenges. As the frequency increases, the wavelength of the SAW decreases, requiring smaller electrode widths and pitch. Fabricating these nanoscale features with high precision becomes increasingly difficult, and any variations in the electrode dimensions can lead to degraded IDT performance. Additionally, at higher frequencies, the IDT's impedance matching becomes more critical, and the device becomes more susceptible to parasitic effects, such as electrode resistance and stray capacitances. These factors can limit the performances of IDT, characterized by its conversion efficiency and bandwidth, directly impacts the modulation efficiency of the device. Although numerical simulations can provide a rough guide about detailed geometry parameters of the IDT to optimize the conversion efficiency, the practical Ohm loss of electrodes and other fabrication imperfection prevent us from predicting the efficiency of the IDT directly. 

\begin{figure}[ht]
	\centering
	\includegraphics[width=1\linewidth]{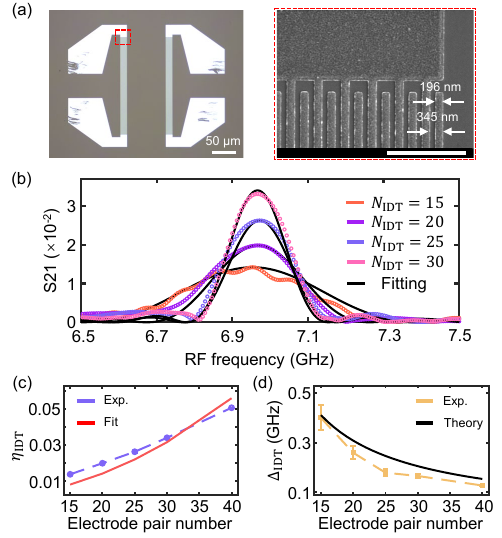}
	\caption{(a) The optical micrograph and the SEM image of IDT pair. The length of each IDT is \SI{200}{\micro\meter}, and the number of electrode pair is 20. The distance between two IDTs is \SI{80}{\micro\meter}.  From the SEM image of IDT, the width of each electrode is about \SI{196}{\nano\meter}, and the distance between two adjacent electrode is about \SI{345}{\nano\meter}. The scale bar of the SEM image is \SI{2}{\micro\meter}. (b) The S21 transmission spectra of several IDT pairs with different electrode pair number. (c) and (d) The experimental results (Theoretical results) of conversion efficiency $\eta_{\mathrm{IDT}}$ (RF to SAW) and the bandwidth $\Delta_{\mathrm{IDT}}$ versus electrode pair number $N_{\mathrm{IDT}}$. 
	}
	\label{fig2}
\end{figure}

We firstly experimentally investigated the relationship between the number of electrode pairs and the IDT's performance metrics. For application in AOM, we only need one IDT but we could not precisely characterize the SAW excitation efficiency of the IDT through the RF reflection spectrum (S11). Therefore, we fabricated several pairs of IDT with varying numbers of electrode finger pairs ($N_{\mathrm{IDT}}$), with the width of each metal electrode is fixed at \SI{196}{\nano\meter}, and the length of each metal electrode is fixed at \SI{200}{\micro\meter}. For example, as shown in Fig.~\ref{fig2}(a) are the optical micrograph of a pair of IDT and SEM image of the local area of a IDT, with $N_{\mathrm{IDT}}= 20$. The pairs of IDT are made by idential IDTs without an optical waveguide, one IDT is used for generating the SAW and the other IDT is used for converting the SAW back to a RF signal after propagation along the LN surface with a distance of \SI{80}{\micro\meter}. 

The S21 transmission spectrum of the IDT pairs are characterized by using a vector network analyzer (Keysight, N5071C). Fig.~\ref{fig2}(b) displays the S21 transmission spectra for different IDTs with varying $N_{\mathrm{IDT}}$. The amplitude transmission spectrum of the IDT pair (S21) with respect to the RF frequency ($f$) follows
\begin{align}
	T_{\mathrm{RF}}(f)& = \eta_{\mathrm{IDT},0} \cdot N_{\mathrm{IDT}}^2 \cdot \mathrm{sinc}^2 \left(\pi N_{\mathrm{IDT}} \frac{f-f_0}{f_0} \right).
\end{align}
Here, $ \eta_{\mathrm{IDT},0}$ is a coefficient corresponding to the energy conversion efficiency between the RF and SAW by a IDT made by one pair of electrode pairs, and $f_0$ is the central frequency of the IDTs. Since the two IDTs are identical, we could estimate the amplitude conversion efficiency of a single IDT as $\sqrt{T_{\mathrm{RF}}(f)}$, and consequently the energy conversion efficiency of a single IDT is $\eta_{\mathrm{IDT}}(f)=T_{\mathrm{IDT}}(f)$. By analyzing the S21 spectra, we extracted the peak energy conversion efficiency $\eta_{\mathrm{IDT}}=\eta_{\mathrm{IDT}}(f_0)$ and the 3\,dB-bandwidth $\Delta_{\mathrm{IDT}}$, as plotted by dots in Fig.~\ref{fig2}(c) and (d). According to the transmission spectra, we can derive the energy efficiency 
\begin{equation}
	\eta_{\mathrm{IDT}}=\eta_{\mathrm{IDT},0}\cdot  N^2_{\mathrm{IDT}},
\end{equation}
and the bandwidth 
\begin{equation}
	\Delta_{\mathrm{IDT}}= 0.886 \cdot \frac{f_0}{N_{\mathrm{IDT}}}.
\end{equation}
By fitting the experimental results in Fig.~\ref{fig2}(c), we obtain $\eta_{\mathrm{IDT},0}=(3.5\pm0.75)\times10^{-5}$. As expected, the IDT becomes more efficient in converting the RF signal to SAWs when the number of electrode pairs increases. The experimental conversion efficiency exhibits an approximate linear increasing with the number of electrode pairs, showing a deviation from the theoretical prediction, might be attributed to the Ohm losses and fabrication imperfections. The theory predicts a reduction of bandwith when increasing $N_{\mathrm{IDT}}$, as shown by the line in Fig.~\ref{fig2}(d), showing good agreement with the experimental results. To achieve a high efficient on-chip AOM, we prioritized conversion efficiency and set the number of electrode pairs to $N_{\mathrm{IDT}}=40$. With this configuration, the experimental conversion efficiency reaches $\eta_{\mathrm{IDT}}= 5\%$, and the corresponding bandwidth is $\Delta_{\mathrm{IDT}}=\SI{128}{MHz}$.

\section{\label{sec:level4}Acousto-optics modulation}
\begin{figure}[ht]
	\centering
	\includegraphics[width=\linewidth]{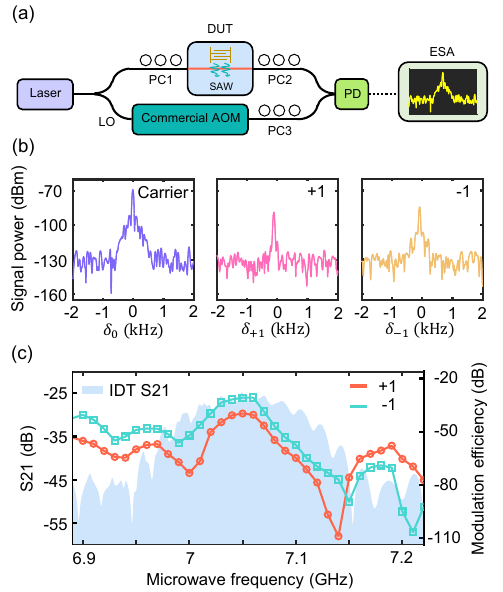}
	\caption{(a) Schematic diagram of the experiment setup. The heterodyne beat measurement is used to measure the modulation efficiency of each sideband of our device. Laser : \SI{780}{\nano\meter}. PC : Polarization Controller. Commercial AOM : bulk acousto-optic modulation, the center frequency is \SI{110}{MHz}. It's used to tune the frequency of the local oscillator.  DUT : device under test. PD : a high-speed photodetector. ESA : the electric spectrum analyzer. (b) These graphs are relative optical amplitude of carrier, the positive first-order sideband and the negative firest-order sideband, respectively. The resolution bandwidth (RBW) is \SI{39}{Hz}, corresponding to the noise level which is about \SI{-128}{dBm}. (c) The experimental results of the S21 transmission spectra of the on-chip IDT pair and modulation efficiency of each sideband versus the microwave frequency. The microwave's drive power is about \SI{30}{dBm}.}
	\label{fig3}
\end{figure}

The performance of our AOM device is characterized by employing the experimental setup illustrated in Fig.~\ref{fig3}(a).  A continuous-wave semiconductor laser (Waviecle) operating at \SI{780}{\nano\meter} is split into two beams, with one beam serving as the optical probe signal, which is coupled into and out from the optical waveguide on the chip using end-face coupling with two lensed fibers, and the other serving as a local oscillator (LO) for heterodyne detection. The phase modulation induced by the AOM generates several frequency sidebands separated by the drive frequency ($\Omega$) applied to the AOM. The power of the carrier ($0$-th order sideband)  and the $\pm1$st-order sidebands are measured through heterodyne beating with the LO, whose frequency is shifted by \SI{110}{MHz} using a commercial AOM (Chongqing Smart Sci$\&$Tech, SGT-110MHz-780-1 TA). The beating signals are detected by a high-speed photodetector (Newport 818-BB-45F) and analyzed with a RF spectrum analyzer (Rohde \& Schwarz, FSP13). 

Fig.~\ref{fig3}(b) shows a representative spectrum obtained with an $8\,\mathrm{mW}$ input optical carrier power and a $30\,\mathrm{dBm}$ RF drive power at $\Omega=7\,\mathrm{GHz}$. The device has a on-chip waveguide length of $3\,\mathrm{mm}$, yielding a carrier transmission efficiency of $0.8\%$. Considering the typical side-coupling efficiency of $20\%$ in our setup, we estimate the corresponding intrinsic optical quality factors of microring resonator made by our waveguides approaches $10^4$. The spectrum exhibits three distinct peaks corresponding to the carrier (blue), $+1$ sideband (red) and $-1$ sideband (yellow), centered at \SI{110}{MHz}, \SI{7.11}{GHz}, and \SI{6.89}{GHz}, respectively. Here,  frequencies $\delta_0$, the $\delta_1$,  $\delta_{-1}$ denotes the relative frequency detuning with respect to the center frequencies. The noise floor of the RF spectrum is \SI{-128}{dBm},  determined by the resolution bandwidth of \SI{39}{Hz} and a spectrum span of \SI{4}{kHz}. 

To gain deep insights into the AOM process, we derived the interaction between the SAW and optical waveguide modes, and numerically investigated the interaction strength and the efficiency of the sideband generation. According previous studies~\cite{Sarabalis2020,qinqing}, the SAW could modulate the permittivity of the materials as $ \Delta \varepsilon (\mathbf{r},t) =\varepsilon_0 \Delta \varepsilon_r (x,y)\cos{(2\pi \Omega t - q_z z)}$, where $\varepsilon_0$ is vacuum permittivity, $\Delta\varepsilon_r(x,y)$ is the change of relative permittivity at the cross-section of the waveguide due to the moving boundary, the photoelastic and the electro-optic effects by the acoustic wave, with $q_z$ is the wavevector of acoustic wave along the propagation direction of optical modes ($z$-axis) and $\Omega$ is acoustic frequency. For the design that the SAW propagates perpendicular with respect to the waveguide, we have $q_z \approx 0$. Due to the wavevector mismatching between different optical modes, we only consider the modulation of the effective index of optical modes as 
\begin{align}
	\delta n (t) =  \frac{n}{2}\frac{\iint dx dy|E(x,y)|^2\Delta \varepsilon_r (x,y)}{\iint dx dy|E(x,y)|^2\varepsilon_r}\cos{(2\pi \Omega t+\phi)},
\end{align}
where $n$ is the effective refractive index and $E(x,y)$ is the electric field distribution of the optical waveguide mode and $\phi$ is the initial phase for the interaction. The modulation leads to the a time-dependent phase modulation of the transmitted light as $\Phi(t)=\delta n(t)2\pi L/\lambda=gL\cos{(2\pi \Omega t+\phi)}$, with $g$ is the interaction strength, $L$ is the length of interaction region and $\lambda$ is the vacuum wavelength of the signal light. According to the Jacobi-Anger expansion, the amplitude transmittance of light can be expressed as
\begin{equation}
	T_{\mathrm{opt}}(t)=e^{-i\omega t-i\Phi(t)}=e^{-i \omega t}\sum_{m=-\infty}^{+\infty} J_m(gL)e^{-i 2\pi m\Omega t},
	\label{sidebands}
\end{equation}
with $J_m$ denotes the $m$-th order Bessel function of the first kind. Therefore, the intensity of $m$-th sideband $\propto J_{m}^2(gL)$. By finite-element simulations, the SAW-induced strain distribution and the optical mode profiles can be numerically solved, and then we obtain the corresponding normalized interaction strength $G=131.4\,\mathrm{(rad/m)/\sqrt{W/m}}$, with $g=G\sqrt{P_{\mathrm{SAW}}/L}$ and $P_{\mathrm{SAW}}$ is the SAW's power. The SAW's power can be calculated based on the power of the incident RF signal. The relationship between the power of SAW and the power of the incident RF signal is given by $P_{\mathrm{SAW}} = P_{\mathrm{RF}}\cdot \eta_{\mathrm{IDT}}$, where $\eta_{\mathrm{IDT}}$ represents the conversion efficiency of the IDT in converting the RF signal into the SAW. For $P_{\mathrm{RF}}=1\,\mathrm{W}$ ($30\,\mathrm{dBm}$) and $L=\SI{200}{\micro\meter}$, we predict an the corresponding $gL=0.41\,\mathrm{rad}$ for an IDT efficiency $\eta_{\mathrm{IDT}}=4.8\%$, indicating that the energy ratio between $\pm1$ sidebands and the carrier is $4\%$.

Fig.~\ref{fig3}(c) displays the experimentally extracted modulation efficiency for the $+1$ (red line) and $-1$ (green line) sidebands, along with the IDT S21 transmission spectrum (blue line). The trend in the modulation efficiency spectrum is consistent with the trend in the IDT S21 transmission spectrum when the RF frequency is close to the center frequency of the IDT. The maximum modulation efficiency achieved for the $-1$ sideband is about \SI{-30}{dB}, while the theoretical modulation efficiency is \SI{-14}{dB}. The discrepancy between the measured and theoretical modulation efficiencies can be attributed to several factors, such as the fabrication imperfections, non-optimal overlap between the optical mode and SAW-induced refractive index modulation, and the presence of acoustic reflections or scattering. 

\section{\label{sec:level5}Sideband asymmetry}
One intriguing aspect of our visible AOM device is the asymmetry observed in the power of the two modulation sidebands, as shown in in Fig.~\ref{fig3}(c). The asymmetry deviates from the predictions of the above theory of phase modulation on a single waveguide mode, according to Eq.~(\ref{sidebands}) and $J_{-m}(x)=(-1)^m J_m(x)$. To verify this asymmetry, we conducted experiments by varying the polarization of the LO light in the heterodyne measurements while keeping all other experimental parameters fixed.  In our setup, the optical fibers used for coupling on-chip waveguide are not polarization-maintaining fiber, and the optical polarization of the input to and output from the chip cannot be determined. In our experiments, we secured the optical fiber onto the optical platform with tape to minimize the polarization state fluctuations of the light field within the fiber during the experiment.
\begin{figure}[ht]
	\centering
	\includegraphics[width=\linewidth]{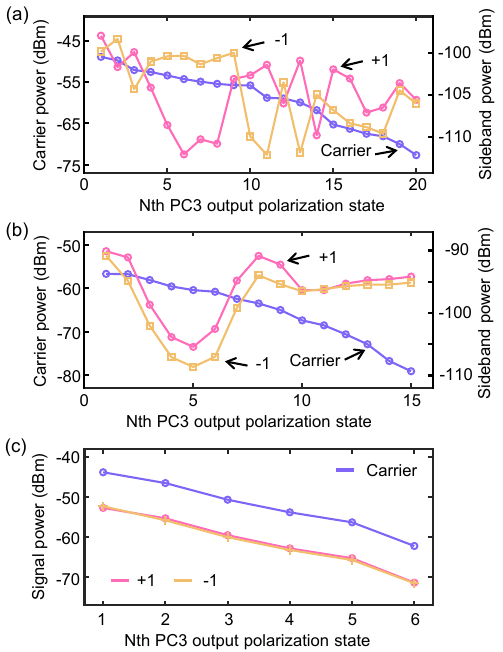}
	\caption{Polarization-dependent sideband asymmetry in the AOM. (a) and (b) Measured power of the carrier (blue), $+1$ sideband (red), and $-1$ sideband (yellow) for different polarization states of the local oscillator light in the heterodyne detection setup. The LO polarization is varied by adjusting polarization controller (PC3), and the results in (a) and (b) are obtained with different input polarizations by adjusting the PC1 to minimize and maximize the carrier transmission, respectively. (c) Control experiment using a commercial fiber-coupled electro-optic modulator (EOM), with the input polarization is adjusted for maximal transmission.
	}
	\label{fig4}
\end{figure}

By adjusting the polarization controller (PC3) in the LO path, as shown in Fig.~\ref{fig3}(a), we recorded the changes in the power of carrier ($I_0$) and sidebands ($I_{\pm1}$) for each LO polarization states. The data set are re-ordered by the descending order of $I_0$ and the results are plotted in Fig.~\ref{fig4}(a) and (b), corresponding to different input polarization achieved by adjusting PC1 to vary the carrier transmittance. We observed that the trend of the sideband power $I_{+1}$ and $I_{-1}$ differ significantly from that of the carrier power $I_{0}$, indicating that the polarization of the sidebands differs from the carrier. To confirm that this behavior is unique to our on-chip AOM, we performed a control experiment using a commercial fiber electro-optic phase modulator (EOM) with the same RF modulation, as shown in Fig.~\ref{fig4}(c). The results from the fiber EOM exhibit symmetric sideband power and consistent polarization behavior across the carrier and sidebands, confirming that our measurement setup is stable. This comparison highlights the distinct polarization-dependent sideband asymmetry in our on-chip AOM, which cannot be explained by the conventional theory of phase modulation on a single waveguide mode.

\begin{figure}[ht]
	\centering
	\includegraphics[width=\linewidth]{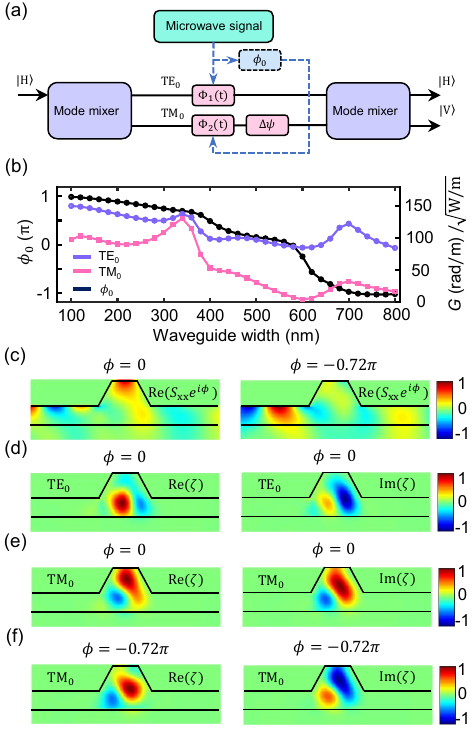}
	\caption{ Multimode acousto-optic interaction and phase-delay effects. (a) Schematic of the proposed multimode acousto-optic modulation model, incorporating the fundamental $\mathrm{TE}_{0}$ and $\mathrm{TM}_{0}$ modes. The mode mixers account for the excitation and interference of the two modes, while the acousto-optic interaction induces phase modulation with a relative delay ($\phi_0$) between the modes. (b) Numerically calculated acousto-optic coupling strength for $\mathrm{TE}_{0}$ (blue) and $\mathrm{TM}_{0}$ (red) as a function of waveguide width. The black curve shows the phase difference between the coupling strengths. (c) Profiles of the strain field ($S_{\mathrm{XX}}$) for phase delays of 0 and $-0.72\pi$, respectively. (d)-(e) Real and imaginary parts of the refractive index modulation $\zeta$ for the $\mathrm{TE}_{0}$ and $\mathrm{TM}_{0}$ modes, respectively, at zero phase dealy. (f) Real and imaginary parts of $\zeta$ for $\mathrm{TM}_{0}$ mode at a phase delay for $\phi=-0.72\pi$.
	}
	\label{fig5}
\end{figure}

To explain the observed sideband asymmetry, we propose a multimode AOM model that incorporates the fundamental transverse-electric ($\mathrm{TE}_0$) and transverse-magnetic ($\mathrm{TM}_0$) modes in the interaction region, as illustrated in Fig.~\ref{fig5}(a). It is important to note that the waveguide in the interaction region is designed to be single-mode, with higher-order TE or TM modes being cut-off. The model includes two mode mixers, which account for the potential excitation and interference of the $\mathrm{TE}_0$ and $\mathrm{TM}_0$  modes. The mode mixer is characterized by a transfer matrix $T(k)$, which is given by  
\begin{equation}
	T(k)= \begin{bmatrix} \sqrt{k} & \mathrm{i} \sqrt{1-k} \\ \mathrm{i} \sqrt{1-k} & \sqrt{k} \end{bmatrix},
\end{equation}
where $k$ represents the power splitting ratio between the TE and TM modes. This model is general and can be reduced to a simple single-mode AOM model by setting both the power splitting ratios $k_1$  and $k_2$ of the input and output mode mixers to $1$. The acousto-optic interaction leads to the phase modulation of optical modes independently, which can be expressed as
\begin{equation}
	M = \begin{bmatrix} e^{\mathrm{i}\Phi_1(t)} & 0 \\ 0 & e^{\mathrm{i}\Phi_2(t)} \end{bmatrix} .
\end{equation}
Here, $ \Phi_1(t) = \Phi_{\mathrm{TE}_0} \cos(2\pi \Omega t )$ and $\Phi_2(t)= \Phi_{\mathrm{TM}_0} \cos(2\pi \Omega t + \phi_{0}) + \Delta\psi $ denotes the phase modulation function for  $\mathrm{TE}_0$ and the $\mathrm{TM}_0$ modes, respectively, $\Phi_{\mathrm{TE}_0} $($\Phi_{\mathrm{TM}_0} $) represents the modulation depths, and $\Delta\psi $ is the relative phase shift between the two modes due to the differences of optical mode effective indices. In particular, $\phi_0 $ is the initial phase difference for the RF signal, which is determined by the independent acousto-optics interaction of two modes. 

Then, the overall transfer function of the multimode AOM can be expressed as:
\begin{equation}
	\begin{bmatrix} E_{\mathrm{out},1}\\ E_{\mathrm{out},2}  \end{bmatrix}	
	= T(k_2) M T(k_1)
	\begin{bmatrix} E_{\mathrm{in},1}\\ E_{\mathrm{in},2}  \end{bmatrix},	
\end{equation}
where $E_{\mathrm{in(out)},1}$ and $E_{\mathrm{in(out)},2}$ are the input(output) field amplitudes of  $\mathrm{TE}_0$ and $\mathrm{TM}_0$ modes, respectively. Assuming  that the incident laser is initially in the TE polarization, i.e., $E_{\mathrm{in},1} = E_0 e^(\mathrm{i}\omega t )$,  and  $E_{\mathrm{in},2} = 0$, we can derive the output field amplitude for the TE polarization as:
\begin{align}
	&E= \sqrt{k_1 k_2} e^{\mathrm{i}(\omega t  + \Phi_1(t))} - \sqrt{(1-k_1)(1-k_2)} e^{\mathrm{i} (\omega t + \Phi_2(t))} \nonumber \\ 
	& = \sqrt{k_1 k_2} e^{\mathrm{i}\omega t } \sum_{n = -\infty}^{\infty} {\mathrm{i}}^n J_n(\Phi_{\mathrm{TE}_0}) e^{\mathrm{i} n 2\pi \Omega t }  \nonumber \\
	&- \sqrt{(1-k_1)(1-k_2)} e^{\mathrm{i}(\omega t  + \Delta\psi)} \sum_{n = -\infty}^{\infty} {\mathrm{i}}^n J_n(\Phi_{\mathrm{TM}_0}) e^{\mathrm{i}(n 2\pi \Omega t + n \phi_0)}.\nonumber
\end{align}
The corresponding power of carrier and sidebands can be calculated as
\begin{align}
	&P_{+1} = k_1 k_2 J^2_{+1}(\Phi_{\mathrm{TE}_0}) + (1-k_1)(1-k_2)J^2_{+1}(\Phi_{\mathrm{TM}_0}) \nonumber  \\ 
	& -2\sqrt{k_1 k_2 (1-k_1)(1-k_2)}  J_{+1}(\Phi_{\mathrm{TE}_0}) J_{+1}(\Phi_{\mathrm{TM}_0}) \cos(\Delta\psi + \phi_0 ) \nonumber \\
	& P_{-1} =  k_1 k_2 J^2_{-1}(\Phi_{\mathrm{TE}_0}) + (1-k_1)(1-k_2)J^2_{-1}(\Phi_{\mathrm{TM}_0}) \nonumber \\ 
	& -2\sqrt{k_1 k_2 (1-k_1)(1-k_2)}  J_{-1}(\Phi_{\mathrm{TE}_0}) J_{-1}(\Phi_{\mathrm{TM}_0}) \cos(\Delta\psi - \phi_0) \nonumber\\
	& P_0 =  k_1 k_2 J^2_{0}(\Phi_{\mathrm{TE}_0}) + (1-k_1)(1-k_2)J^2_{0}(\Phi_{\mathrm{TM}_0}) \nonumber \\ 
	& -2\sqrt{k_1 k_2 (1-k_1)(1-k_2)}  J_{0}(\Phi_{\mathrm{TE}_0}) J_{0}(\Phi_{\mathrm{TM}_0}) \cos(\Delta\psi ). \nonumber
\end{align}

By analyzing these expressions, we find that the sideband asymmetry is forbidden if the modulation phase difference between the two modes is zero, i.e., $\phi_0=0$, regardless of the values of the other parameters $k_1$, $k_2$, $\Delta\psi$, $\Phi_{\mathrm{TE}_0}$, and $\Phi_{\mathrm{TM}_0}$. This result highlights the crucial role of the modulation phase difference in enabling the observed sideband asymmetry. It is worth noting that this multimode AOM model also encompasses the potential mechanism where the input mode is converted into sidebands with orthogonal polarization. For example, the TM mode might be the sidebands generated by the input of pure TE mode. This situation is mathematically equivalent to the case where the input is mixed polarization, but only TM mode phase modulation is allows as $\Phi_{\mathrm{TE}_0} =0$ and $\Phi_{\mathrm{TM}_0}\neq0$). However, for such configuration, there should be no modulation phase difference ($\phi_0=0$) and the output should not exhibit asymmetric sidebands. 

Therefore, the experimental results and our theoretical analysis suggests that two key features of our AOM: (1) two orthogonally-polarized modes are excited in our device, and (2) the modulations of these modes have a relative phase difference. The excitation of orthogonal polarization modes can be attributed to unintentional modal coupling due to waveguide tapering, surface roughness, or polarization mixing in the input and output fibers. However, the modulation phase difference is non-trivial, as the SAW is excited by a simple, uniform IDT structure. We propose that this phase difference arises from the propagation delay of the SAW as it interacts with the strongly localized TE and TM modes. Because the difference in the transverse lateral field distributions of the two optics modes, their optimal overlap with the propagating SAW occurs at different moment. In particular, this effect becomes significant when the wavelength of the SAW is comparable with the waveguide width.

To support this hypothesis, we performed systematic numerical simulations of the interaction between the SAW and the two polarized guided modes. We calculated the phase and amplitude of $G_{\mathrm{TE}}$ and $G_{\mathrm{TM}}$ in Fig.~\ref{fig5}(b) for TE and TM modes, which shows a non-trivial phase delay, which is $\phi_0= \arg(G_{\mathrm{TM}}) -\arg(G_{\mathrm{TE}}) = 0.72\pi$, $0.49\pi$, and $0.16\pi$ for different waveguide width $w=320$, $400$, and $500\mathrm{nm}$. We note that the absolute value $|G_{\mathrm{TE}}|$ and $|G_{\mathrm{TM}}|$ simultaneously reach peak values at certain $w$ due to the constructive interference of SAW in the wedge waveguide~\cite{qinqing}. Specifically, we present the acoustic strain field $S_{\mathrm{XX}}$ and refractive index modulation $\zeta(x,y)=|E(x,y)|^2\Delta \varepsilon_r (x,y)$ profiles at the cross-section of the waveguide ($w=320\,\mathrm{nm}$) in Fig.~\ref{fig5}(c)-(f). The simulation results confirm that the distinct mode profiles and modulation tensors of the material lead to a phase delay between the interactions. Fig.~\ref{fig5}(c) shows the distribution of stain component $S_{\mathrm{XX}}$ at different initial phase. Fig.~\ref{fig5}(d) and (e) are the modulation profiles ($\zeta$) corresponding to the same initial phase of the strain for $\mathrm{TE}_0$ mode and $\mathrm{TM}_0$ mode, respectively. Comparing the real part $\mathrm{Re}(\zeta)$ and imaginary part $\mathrm{Im}(\zeta)$ of $\mathrm{TE}_0$ mode and $\mathrm{TM}_0$ mode, we confirm that it's not synchronous for phase modulation for $\mathrm{TE}_0$ mode and $\mathrm{TM}_0$ mode. Interestingly, if we apply a phase delay ($-0.72\pi$) on the every strain component, as is shown in Fig.~\ref{fig5}(f), the modulation profiles ($\zeta$)  of $\mathrm{TE}_0$ mode and $\mathrm{TM}_0$ mode become similar. 

\begin{figure}[ht]
	\centering
	\includegraphics[width=\linewidth]{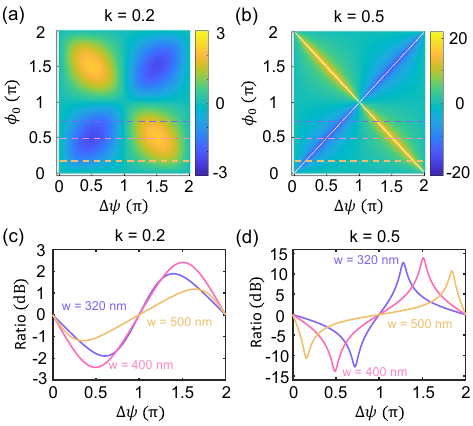}
	\caption{(a) and (b) Calculated sideband power ratio as a function of the propagation-induced phase shift ($\Delta\psi$) and the acoustic-optic interaction phase delay $\phi_0$ for mode mixing ratios $k=0.2$ and $0.5$, respectively. (c) and (d) Sideband power ratio versus $\Delta\psi$ for different waveguide widths (320, 400, and 500\,nm) and mode mixing ratios $k=0.2$ and $0.5$, respectively.
	}
	\label{fig6}
\end{figure}

We choose two different mode mixing splitter ratio ($k = 0.2$ and $k = 0.5$), present the numeric results of the modulation sideband ratio versus propagation-induced relative phase shift $\Delta\psi$ and modulation phase delay $\phi_0$. As shown in Fig.~\ref{fig6}(a) and (b), the mode mixing splitter ratio determine the upper bound and lower bound of the modulation sideband ratio. 

To further illustrate the impact of the modulation phase delay on the sideband asymmetry, we consider the case of $\phi_0=0.72\pi$ as an example. Fig.~\ref{fig6}(c) and (d) show the sideband asymmetry, which is quantified by $10\mathrm{log}_{10}\left(I_{+1}/I_{-1}\right)$, for different values of the propagation-induced relative phase shift $\Delta\psi$ and the mode mixing splitter ratios $k_{1,2} = 0.2$ and $0.5$, respectively. Our analysis reveals several important insights into the behavior of the sideband asymmetry: (1) When the modulation phase delay is an integer multiple of $\pi$, i.e., $\phi_0 = n\pi$ with $n \in \mathbb{Z}$, the sideband asymmetry vanishes. This observation highlights the critical role of the modulation phase delay in enabling the asymmetric sideband generation. (2) The sideband asymmetry reaches its maximum value when the modulation phase delay is an odd multiple of $\pi/2$, i.e., $\phi_0 = n\pi + \pi/2$. In this case, by carefully adjusting the propagation-induced relative phase shift $\Delta\psi$ between the TE and TM modes in the waveguide, it is possible to completely suppress one of the sidebands, leading to a single-sideband modulation. (3) The mode mixing splitter ratios $k_1$ and $k_2$ provide an additional degree of freedom for controlling the sideband power. By designing the device with tunable $k_1$ and $k_2$ values, it becomes possible to simultaneously suppress both the carrier and one of the sidebands, enabling an ultracompact single-sideband modulator based on the multimode AOM.

\section{\label{sec:level6}Discussion}
The realization of an on-chip acousto-optic modulator operating at 7 GHz for visible wavelengths represents a significant milestone in the field of integrated photonics. The compact device footprint of 200 microns, achieved through the novel design of aligning the SAW perpendicular to the optical waveguide, showcases the potential for dense integration and scalability. However, it is important to address potential challenges and limitations of the current work. One issue is the limited transduction efficiency of the current IDT. Future work should focus on a detailed analysis of the loss mechanisms of the IDT and explore strategies for enhancing the transduction efficiency. Additionally, optimizing the device geometry to improve the SAW-optical mode overlap, reducing acoustic losses, and enhancing the coupling efficiency between the optical waveguide and external systems could further boost the modulation efficiency. Exploring advanced IDT designs, such as apodized or chirped structures, could also help tailor the frequency response and bandwidth of the AOM to meet specific application requirements.

The observed sideband asymmetry arises from the intricate interplay between the SAW and the polarized optical modes at the sub-micron scale. Our configuration is in sharp contrast to other single-sideband mode converter, where an angle between the propagation direction of the SAW and the waveguide is carefully designed for the phase matching between the SAW and two optical modes in the waveguide~\cite{Kittlaus2021,Sohn2021}. Our results only provide new insights into the fundamental mechanisms of acousto-optic modulation but also open up exciting possibilities for polarization-dependent modulation and signal processing in integrated photonic devices. From a practical perspective, the realization of a compact, efficient single-sideband modulator has significant implications for various applications, such as the microwave signal processing via optical frequency up-conversion. To fully harness the potential of this multimode AOM for single-sideband modulation, future work should focus on developing strategies for precise control and dynamic tuning of the modulation phase delay, the propagation-induced phase shift, and the mode mixing ratios.

\section{\label{sec:level7}Conclusion}
In conclusion, this work represents a significant step forward in the development of high-frequency acousto-optic modulators for visible wavelengths, showcasing the potential of the lithium niobate on sapphire platform. The interplay between the SAW and the polarized optical modes at the sub-micron scale gives rise to the observed sideband asymmetry, and our findings highlight the rich physics underlying the acousto-optic interaction in our visible AOM device.  With further optimizations and investigations into the rich physics at play, this work paves the way for a new generation of compact, high-performance, and versatile photonic devices.

\section{\label{sec:level8}Acknowledgments}
This work was funded by the National Key R$\&$D Program (Grant Nos.~2021YFF0603701 and 2021YFA1402004), the National Natural Science Foundation of China (Grants 12074067, 92265108, U21A20433, U21A6006, 12104441, 12061131011, and 92265210). This work was also supported by the Fundamental Research Funds for the Central Universities and USTC Research Funds of the Double First-Class Initiative. The numerical calculations in this paper have been done on the supercomputing system in the Supercomputing Center of the University of Science and Technology of China. This work was partially carried out at the USTC Center for Micro and Nanoscale Research and Fabrication.


\bibliographystyle{Zou}
\nocite{*}
\bibliography{AOM_arxiv}

\end{document}